\documentclass{article}

\usepackage[preprint]{neurips_2025}

\usepackage[utf8]{inputenc} 
\usepackage[T1]{fontenc}    
\usepackage{hyperref}       
\usepackage{url}            
\usepackage{booktabs}       
\usepackage{amsfonts}       
\usepackage{nicefrac}       
\usepackage{microtype}      
\usepackage{xcolor}         
\usepackage{graphicx}
\usepackage{amsmath}
\usepackage{amssymb}
\usepackage{amsthm}
\usepackage{multirow}
\usepackage{wrapfig}
\usepackage{caption}

\title{Harnessing the Power of Reinforcement Learning for Language-Model-Based Information Retriever via Query-Document Co-Augmentation}

{
  \author{Jingming Liu$^{1,*}$\quad Yumeng Li$^{1,}$\thanks{Equal contribution.} \quad Wei Shi$^{2,\dagger}$ \quad Yao-Xiang Ding$^{1,}$\thanks{Corresponding authors.} \quad Hui Su$^2$ \quad Kun Zhou$^1$ \\[1.25ex]
$^1$State Key Laboratory of CAD\&CG, Zhejiang University \quad $^2$Meituan Inc. \\[1.25ex]
\texttt{\{jml754457, shiwei1122, dingyx.gm\}@gmail.com, yumeng.li@zju.edu.cn} \\
\texttt{suhui07@meituan.com, kunzhou@acm.org} \\
}}

\begin{document}

\maketitle

\begin{abstract}
Recent studies have proposed leveraging Large Language Models (LLMs) as information retrievers through query rewriting. However, for challenging corpora, we argue that enhancing queries alone is insufficient for robust semantic matching; the LLM should also have sufficient understanding of the corpus by directly handling and augmenting the documents themselves. To this end, we present an LLM-based retriever empowered to augment both user queries and corpus documents, with its policy fully explored via reinforcement learning (RL) and minimal human inductive bias. Notably, we find that simply allowing the LLM to modify documents yields little benefit unless paired with our carefully designed bidirectional RL framework, which enables the LLM to simultaneously learn and collaborate on both query and document augmentation policies. A key technical challenge in realizing such a framework lies in jointly updating both policies during training, where the rewards for the two directions depend on each other, making their entangled reward intractable. Our approach addresses this by introducing a reward sampling strategy and a specifically designed RL algorithm that enables effective training with these sampled rewards. Experimental results demonstrate that our approach significantly enhances LLM-based retrieval performance in both sparse and dense settings, particularly in difficult retrieval domains, and achieves strong cross-benchmark generalization. Our code is released at \url{https://github.com/liujm2001/CoAugRetriever}.

\end{abstract}

\section{Introduction}
Information retrieval (IR)~\citep{baeza1999modern, singhal2001modern} studies the task of finding the best match from a large set of documents based on a given query, and has played a crucial role in recent AI task scenarios, such as retrieval-augmented generation (RAG)~\citep{gao2023retrieval, wang2024multilingual}. Classical IR approaches include sparse retrievals based on TF-IDF~\citep{salton1988term} and BM25~\citep{robertson2009probabilistic}, as well as dense retrievals based on embeddings obtained from pre-trained language models~\citep{xiao2024c}. With well-established retrievers and the emergence of large language models (LLMs), recent works have identified a bottleneck in poor query quality, thus enhancing IR accuracy through query rewriting~\citep{ma2023query, ye2023enhancing, wang2023query2doc, shen2023large, mao2024rafe}. However, retrieval performance still has room for improvement, especially in challenging knowledge domains where accurately retrieving information from a compact corpus is crucial~\citep{dai2024cocktailcomprehensiveinformationretrieval, dai2024bias}.

In this work, we show that, symmetrically to queries, documents can be better managed by allowing the LLM retriever to develop a sufficient understanding of the corpus through direct handling and augmentation of the documents, thereby pulling challenging queries and documents to be more semantically related and better paired for retrieval. We propose an LLM-retriever that simultaneously augments both user queries and documents, with its policy explored purely through reinforcement learning (RL) with minimal human-designed inductive bias.

Notably, we find that simply allowing the LLM to modify documents yields little benefit unless paired with our carefully designed RL framework to perform joint bidirectional training in a single process, which enables the model to cooperate with itself in both query and document enhancement. A key technical challenge in realizing this training process is the significantly enlarged action space, as the final action is the combination of augmentation actions in both directions, making exact reward computation intractable. We propose a reward sampling strategy for this bidirectional training, and design specialized adjustments to enable the RL algorithm to work with our sampled reward, where direct application of state-of-the-art LLM reinforcement learning algorithms fails to handle our task. To integrate with existing LLM RL frameworks, we adopt a batch-unbatch alternating implementation to achieve effective training under our bidirectional setting.

We conducted experiments on challenging IR benchmarks to verify the efficacy of our approach using both sparse and dense retrievers. The results show that our approach successfully tackled the collaborative training challenge and enabled to learn an effective bidirectional augmentation policy, which significantly enhances the performance of both the base model and query enhancement methods. We also provide an analysis of the behavior of the trained policy using our approach, which helps reveal the underlying causes of the improved performance. Furthermore, we observe that the learned LLM-retriever policy achieves desirable cross-benchmark generalization ability, demonstrating that our approach successfully harnesses the power of RL to enable the LLM-retriever to obtain generalizable capabilities in IR through self-exploration.

\begin{figure}
    \centering
    \includegraphics[width=0.85\linewidth]{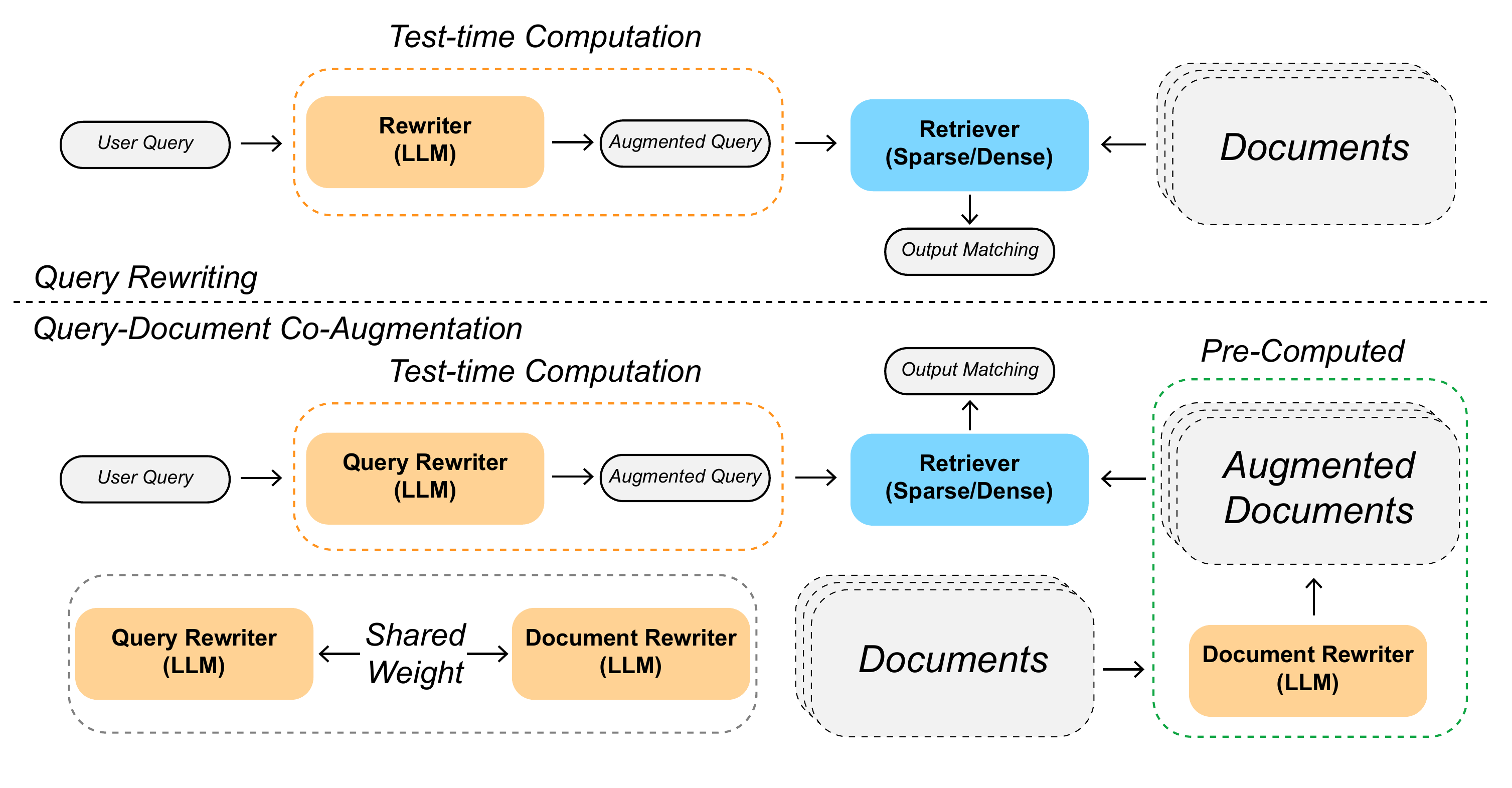}
    \caption{Alongside ordinary test-time query rewriting, we allow the LLM to pre-compute lightweight augmentations for every document, giving the LLM control over both queries and documents. This query–document co-augmentation widens the action space of the LLM, enabling the exploration of more accurate retrieval policies, especially in challenging collections.}
    \label{fig:illustration}
\end{figure}

\section{Related Work}
\label{sec:related_work}

With the wide adoption of LLM systems, information retrieval (IR) systems are gaining increasing importance through retrieval-augmented generation (RAG), reducing hallucination and enriching factual knowledge~\citep{mallen2023llm_memorization,shi2023replug,chen2017reading,lee2019latent,guu2020retrieval,lewis2020rag,lazaridou2022internet,asai2024selfrag}. IR systems also benefit from LLMs to increase their retrieval accuracy by rewriting queries~\citep{ma2023query}, as the queries might not provide sufficient information that can be faithfully related to the document, hampering IR accuracy~\citep{dageville2016snowflake, belkin1982ask}. In this work, we focus on enhancing the ability of LLMs to improve IR accuracy using reinforcement learning. We will first discuss retrieval augmentation methods, then discuss methods for enhancing LLMs in IR tasks.

\subsection{Retrieval Augmentation}
Previous methods mainly enhance information retrieval tasks by developing better retrievers~\citep{chen2017reading,karpukhin-etal-2020-dense}, enhancing retrievers and readers together~\citep{karpukhin-etal-2020-dense,lewis2020rag,emdr22021SachanRHDY21,lee-etal-2022-need,jiang-etal-2022-reatt}, or by enhancing queries with external knowledge~\citep{zesch2007analyzing,zesch-etal-2008-extracting,syed2010wikitology,Dalton2014EntityQF,Xu2009QueryDP,Meij2010ConceptualLM,Xiong2015QueryEW} and relevant content~\citep{abdul2004umass, metzler2005markov, metzler2007latent}.

In the era of LLMs, as sparse retrievers are well-established and tuning dense retrievers requires a lot of data to reduce overfitting~\citep{ma2023query}, recent methods mainly focus on leveraging or improving the ability of LLMs in information retrieval~\citep{trivedi2023interleavingretrievalchainofthoughtreasoning,yao2023reactsynergizingreasoningacting,khattab2022demonstrate,press2022measuring}. Studies have revealed that, pre-trained on large corpora, LLMs without fine-tuning already serve as powerful query optimizers~\citep{shen2023large,wang2023query2doc,brown2020language,Touvron2023Llama2O}.

\subsection{Reinforcement Learning for Enhancing LLMs in IR}
With the success of RLHF in aligning LLMs with human preferences~\citep{Christiano2017DeepRL,Stiennon2020LearningTS,Ouyang2022TrainingLM}, reinforcement learning has emerged as a principled approach for enhancing LLMs, such as PPO~\citep{Schulman2017ProximalPO}, as well as recent methods including GRPO~\citep{deepseekai2025deepseekr1incentivizingreasoningcapability} and REINFORCE++~\citep{Hu2025REINFORCEAS}, which have demonstrated significant performance gains in tasks by exploring based on reward beyond preference alignment~\citep{deepseekai2025deepseekr1incentivizingreasoningcapability}.

In the domain of information retrieval, recent methods have also explored utilizing reinforcement learning to improve query augmentation. Some systems leverage reward feedback from the final search or generation results~\citep{ma2023queryrewritingretrievalaugmentedlarge, Fan2024ASO, Zhao2024RetrievalAG}. Another line of work, which is most relevant to us, directly utilizes feedback signals, including a recent work exploring trial-and-error from metric feedback~\citep{hsu2024grounding} and a concurrent work~\citep{jiang2025deepretrieval} utilizing feedback signals to enhance IR tasks and SQL tasks. Our work also focuses on improving the performance of LLMs in the specific IR task. The main difference is that our proposed method enables the LLM to explore a policy that not only augments the query but also manages the document itself simultaneously, which significantly improves performance in challenging IR domains.

\section{Proposed Approach}
\label{sec:approach}
\subsection{Overview}

We propose enhancing a large language model (LLM) via reinforcement learning (RL) to jointly process queries and documents, thereby aligning their word distributions and semantic spaces with the model's internal knowledge for improved retrieval performance. After training, the learned policy can be used to precompute augmented document representations, effectively encoding document knowledge in advance. At inference time, given a user query, the policy augments both the query and performs retrieval over the preprocessed document collection with computational costs comparable to standard query rewriting. Since query and document augmentation are formulated as text processing tasks, they remain independent of the underlying retrieval method, enabling flexible integration with various retrieval modules (e.g., BM25~\citep{robertson2009probabilistic} for sparse retrieval or BGE models~\citep{xiao2024c} for dense retrieval).

\begin{figure}[t]
    \centering
    \includegraphics[width=0.9\linewidth]{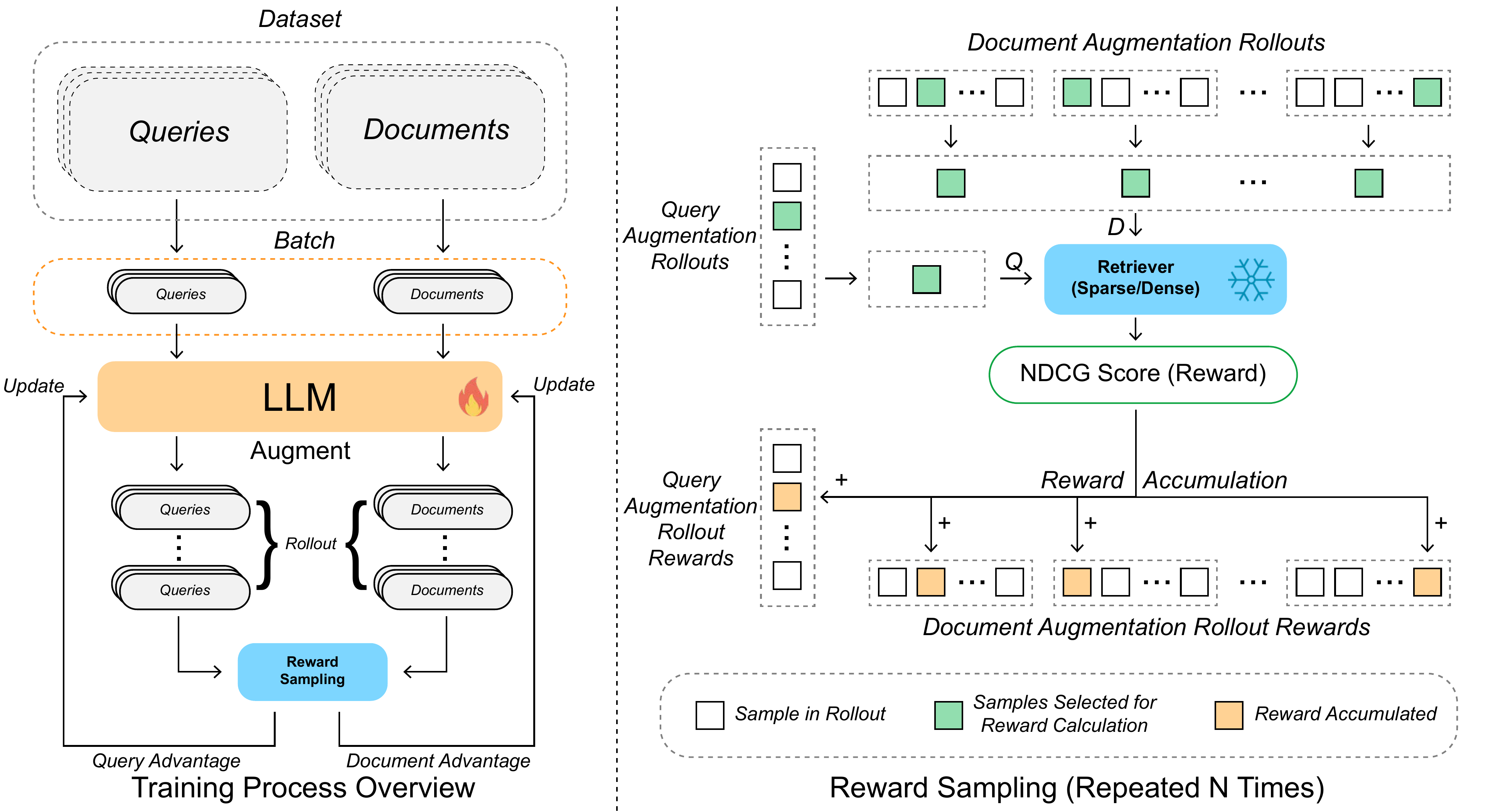}
    \caption{Illustration of the bidirectional RL training process. Left: Overview of the training pipeline. Right: Reward calculation—one sample is randomly selected from the rollouts for each query and related document, its reward is computed, and the reward is accumulated onto the corresponding rollout sample for estimation.}
    \label{fig:training}
\end{figure}

The overview of the training pipeline is illustrated in Fig.~\ref{fig:training}. To enable joint training of query and document augmentation, we introduce a novel query-document composite sampling strategy (Sec.~\ref{query_document_composite_sampling}). This approach organizes queries and their associated documents into the same batch during sampling, effectively reducing the training data scale from the entire document collection to a manageable batch size. For reward computation (Sec.~\ref{within_batch_reward_computation}), we perform retrieval by executing query rollouts within document rollouts for each batch, using the average score of each rollout as the reward signal. Owing to the distinct characteristics of our reward computation, conventional group-wise or batch-wise reward normalization methods, such as those used in GRPO~\citep{deepseekai2025deepseekr1incentivizingreasoningcapability} or REINFORCE++~\citep{hu2025reinforce++}, are not directly applicable. Therefore, we adapt the advantage computation method (Sec.~\ref{taming_reward_var_in_rl}) to better align with our task requirements. Notably, despite introducing more sophisticated sampling, reward computation, and advantage calculation procedures, our approach remains fully compatible with standard LLM RL training pipelines (Sec.~\ref{batch_unbatch_alternating_implementation}). Specifically, rollout inference and backpropagation are performed at the individual text level, while sampling, reward computation, and advantage calculation are conducted at the group level.

\subsection{Query-Document Composite Sampling}
\label{query_document_composite_sampling}

To enable synchronous augmentation training for both queries and documents, a straightforward approach would be to evaluate lexical and semantic alignment by performing retrieval with augmented queries over the augmented document collection. However, executing inference at the scale of the entire document collection during every training step is computationally infeasible. To address this, we modify the sampling strategy to preserve the original batch size while ensuring that each batch contains three essential components: queries, relevant documents, and irrelevant documents, thereby forming a representative mini-dataset.

The sampling procedure is as follows: We first randomly select $q$ queries from the query pool. For each query, we sample $d_{pos}$ documents with positive retrieval scores (i.e., relevant documents). Additionally, we sample $d_{neg}$ documents (irrelevant documents) that have zero retrieval scores with respect to all $q$ queries. This combination of queries, relevant documents, and irrelevant documents constitutes a complete data group. Each batch thus contains $q + q \times d_{pos} + d_{neg}$ distinct texts, which are assigned appropriate system prompts based on their type (query or document) before being processed by the text augmentation model as training data.

\subsection{Within-Batch Reward Computation}
\label{within_batch_reward_computation}

Given the inherent semantic and lexical disparities between user queries and documents, employing a single critic model to evaluate augmented texts originating from different distributions may result in suboptimal assessment quality. Following the methodology of GRPO, we generate $n_{rollout}$ augmented rollouts for each text and compute rewards for each variant at the group level. However, the combinatorial complexity increases exponentially with the number of documents: for a batch containing $q + q \times d_{pos} + d_{neg}$ texts, evaluating all possible configurations would require processing:$q \times n_{rollout} \times n_{rollout}^{d_{pos} + d_{neg}}$ matching pairs, which is exponential in the number of documents. 

We observe that while query rollouts remain independent, each combination of a document’s $n_{rollout}$ augmented rollouts introduces potential variations in similarity rankings during retrieval, which consequently influences reward computation. To address this challenge, we adopt a straightforward multi-sampling strategy applied to the mini-batch of documents.

Specifically, in each sampling iteration, we randomly select one rollout for each document and then evaluate all query rollouts by calculating their NDCG scores~\citep{jarvelin2002cumulated} based on the similarity rankings obtained from retrieval. These scores serve as rewards for the selected combinations of query and document rollouts. The final reward for each rollout is calculated by averaging between all sampling iterations. This sampling method incurs significantly lower computational overhead (less than $1e^{-1} \times $ inference time) while accurately estimating the rewards for each query and document rollout (error < $1e^{-2}$), thereby enabling efficient and effective synchronous training of both query and document augmentation components.

\subsection{Taming Reward Variance in RL}
\label{taming_reward_var_in_rl}
\begin{figure}[t]
    \centering
    \includegraphics[width=0.9\linewidth]{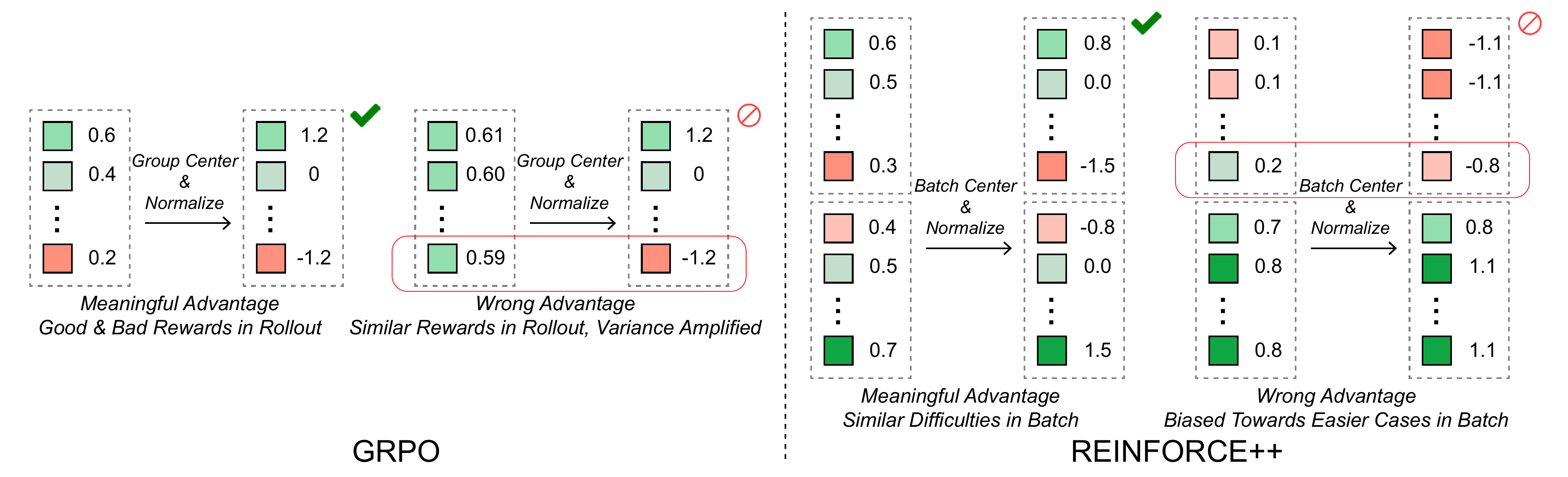}
    \caption{Illustration of the problems identified in our mission when adopting state-of-the-art reinforcement learning algorithms. The unique challenge of our task requires specialized design.}
    \label{fig:sampling_strategy}
\end{figure}
While the sampling strategy serves as an effective reward estimator, it inevitably introduces some variance. However, our within-batch reward computation can easily reduce this variance to the level of $1e^{-5}$, rendering reward fluctuations negligible for policy learning in most cases. In GRPO, the advantage for each rollout is computed as $(r-r_{mean})/r_{std}$. As illustrated in Fig.~\ref{fig:sampling_strategy}, we observe that when all rollouts within a group receive identical rewards, the within-group normalization in GRPO amplifies the originally negligible sampling variance to 1. This leads to random advantage assignments among rollouts, without reflecting meaningful quality differences. Consequently, substantial noise is introduced into the optimization process, ultimately resulting in training failure.

We subsequently investigated REINFORCE++, which modifies the GRPO approach by replacing within-group reward normalization with batch-wide normalization, thereby mitigating noise arising from normalized sampling variance. However, in our experimental setup, significant variations in average scores attributable to differing query difficulties introduce a critical limitation. As shown in Fig.~\ref{fig:sampling_strategy}, REINFORCE++’s batch-wide normalization causes the advantage computation to be dominated by query difficulty rather than the quality of text augmentation. As a result, rollouts from the same sample exhibit minimal variation, often becoming uniformly positive or negative, which leads to poor training performance.

Returning to the fundamentals of reinforcement learning training, our primary concerns regarding advantage computation are numerical stability and discriminative power, rather than standardization. Notably, our implementation of expected NDCG scores as rewards (as detailed in Sec. \ref{within_batch_reward_computation}) naturally satisfies these requirements. These scores maintain stable distributions within the $[0, 1]$ range while effectively capturing variations in augmentation quality.

Our final solution is simple but effective. We remove within-group normalization while retaining within-group centering. This modification successfully eliminates noise arising from identical rewards, while preserving the original reward differentials among group rollouts, resulting in substantial improvements in training performance. Additionally, we apply differential scaling to the advantage scores computed from queries, relevant documents, and irrelevant documents to balance their contributions during training. This adjustment mitigates the risk of gradient domination by the disproportionately large number of irrelevant documents.

To validate these findings, we conducted comprehensive ablation studies comparing various approaches. The respective impacts on training efficacy are discussed in detail in Sec.~\ref{subsec:ablation_studies}.

\begin{figure}
    \centering
    \includegraphics[width=.9\linewidth]{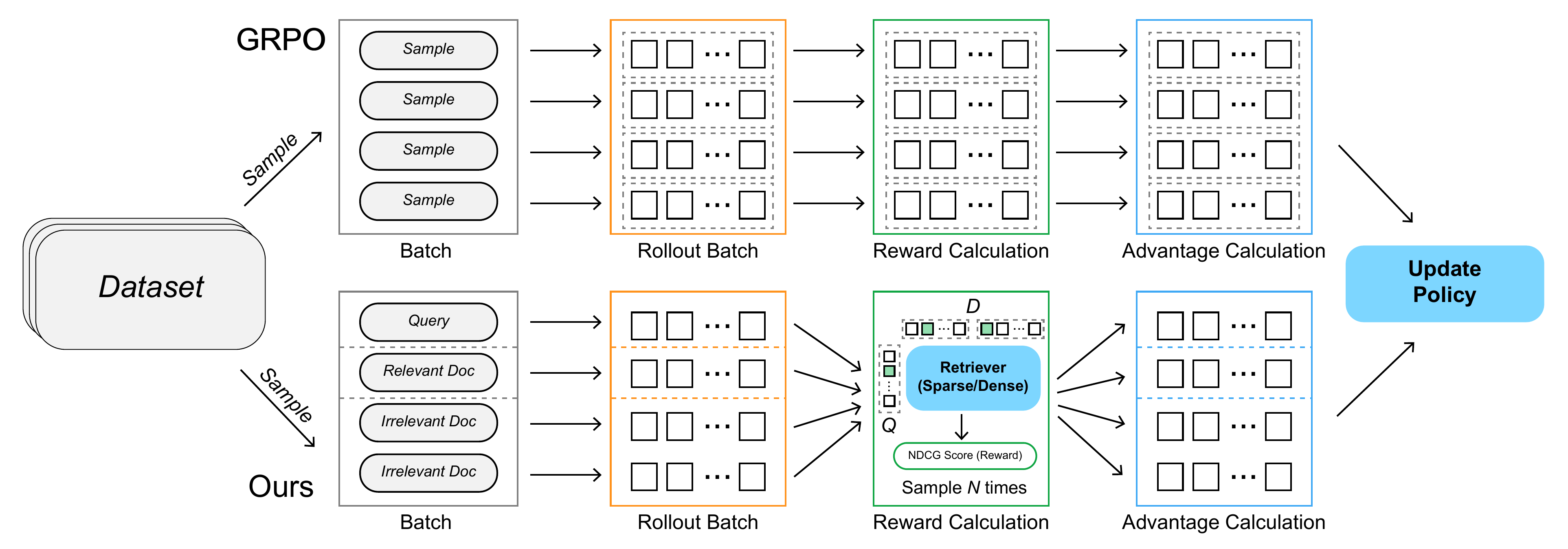}
    \caption{Implementation of our RL training pipeline. The entanglement is introduced by our reward calculation. Thus, we pack queries and documents into a single patch, and perform paralleled sampling for rollout efficiency. We only entangle the contents for reward computation, and revert them to separate samples for advantage computation and policy update.}
    \label{fig:implementation_pipeline}
\end{figure}

\subsection{Batch-Unbatch Alternating Implementation}
\label{batch_unbatch_alternating_implementation}

To align with conventional reinforcement learning training frameworks, we adopt a batched-unbatched alternating paradigm shown in Fig.~\ref{fig:implementation_pipeline}. The core distinction from previous approaches lies in our batch construction, where individual samples are intrinsically interlinked rather than independently distributed.

\textbf{Batch-Level Sampling}. We redesigned the dataset sampling mechanism to generate batches comprising queries, relevant documents, and irrelevant documents, following the methodology described in Sec. \ref{query_document_composite_sampling}. Each text sample is augmented with system prompts according to its source before being assembled into the same batch. This design ensures the integrity of each batch, allowing it to be used independently for training.

\textbf{Sample-Level Inference}. During model rollout operations, batched text samples are logically unbatched into individual samples. At this stage, the distinct sources of the samples (queries or documents) become transparent to the inference engine, enabling parallel processing of all rollouts without mutual interference. Importantly, this phase maintains complete compatibility with existing reinforcement learning training infrastructures.

\textbf{Batch-Level Reward Computation}. All rollouts are reaggregated at the batch level for reward calculation. We group the selected rollouts based on their sources (queries or documents), and then apply the retrieval module to rank them and compute the corresponding rewards. Subsequently, group-wise centralization is performed for each sample to provide advantage estimates for individual rollouts.

\textbf{Sample-Level Parameter Update}. After calculating the advantages, the text samples are unbatched once again. Although processed in batch form, policy gradient losses are individually computed for each rollout using their respective advantage values, and the model parameters are then updated accordingly. This step is fully compatible with existing reinforcement learning training frameworks and does not require additional modifications.

\section{Experiments}

\begin{table*}[t]
    \centering
        \caption{NDCG@10 Performance under sparse (BM25) and dense (BGE-base-en-v1.5) retrieval. ``Ours-NF, Ours-SF, Ours-FQ'' indicate the model utilizing our approaches trained on NFCorpus, SciFact, and FiQA-2018. The best results are bolded, and other top-three results are underlined.}
    \resizebox{.5\linewidth}{!}{
        \begin{tabular}{ccccc}
        \toprule
        \multicolumn{2}{c}{Settings} & NFCorpus & SciFact & FiQA-2018 \\ \midrule

         \multirow{2}{*}{Base Retriever}& BM25& 0.343& 0.691& 0.254 \\ 
                            & BGE-base-en-v1.5& 0.368& 0.738& \underline{0.391} \\ \midrule
         \multirow{2}{*}{Qwen2.5-7B    }& BM25& 0.363& 0.696& 0.258 \\ 
                            & BGE-base-en-v1.5& \underline{0.371}& \underline{0.746}& \underline{0.383} \\ \midrule
         \multirow{2}{*}{Ours-NF       }& BM25& \bf 0.403& \underline{0.741}& \underline{0.272} \\ 
                            & BGE-base-en-v1.5& \bf 0.384& \underline{0.743}& 0.371 \\ \midrule
         \multirow{2}{*}{Ours-SF       }& BM25& \underline{0.378}& \bf 0.748& \underline{0.268} \\ 
                            & BGE-base-en-v1.5& \underline{0.378}& \bf 0.753& 0.376 \\ \midrule
         \multirow{2}{*}{Ours-FQ       }& BM25& \underline{0.386}& \underline{0.744}& \bf 0.328 \\ 
                            & BGE-base-en-v1.5& 0.369 & \underline{0.743}& \bf 0.395 \\ 
\bottomrule
    \end{tabular}}
    \label{tab:quantitative}
\end{table*}

\begin{table*}[t]
    \centering
        \caption{NDCG@10 performance of query-only and doc-only ablation studies on NFCorpus. Base-Q and Base-D represent using the base model (Qwen2.5-7B) to enhance only queries or only documents, respectively. RL-Q and RL-D refer to models trained using RL for query-only augmentation and document-only augmentation, respectively. The plus sign (+) indicates that the methods are used jointly. RL-QD refers to collaborative training for bidirectional augmentation (our proposed method).}
    \resizebox{1\linewidth}{!}{
        \begin{tabular}{ccccccccc}
        \toprule
 Settings &  Base Retriever & Base-Q & Base-D & Base-Q + Base-D & RL-Q & RL-D & RL-Q + RL-D & RL-QD (ours) \\
 \midrule
BM25 & 0.343 & 0.357 & 0.356 & 0.363 & 0.381 & 0.372 & 0.388 & \bf 0.403 \\
BGE-base-en-v1.5 & 0.368 & 0.377 & 0.364 & 0.371 & 0.379 & 0.373 & 0.372 & \bf 0.384 \\
\bottomrule
    \end{tabular}}
    \label{tab:ablation_on_nfcorpus_ndcg}
\end{table*}

\begin{table}[t]
    \centering

    \begin{minipage}{0.4\linewidth}
        \centering
        \captionof{table}{Cross Entropy of query-only and doc-only Ablation Studies on NFCorpus.}
    \resizebox{1\linewidth}{!}{
        \begin{tabular}{clr}
        \toprule
        \multicolumn{2}{c}{Settings} & H(Q, D)\\ \midrule
         \multirow{2}{*}{Qwen2.5-7B    }& BM25& 10.318 \\ 
                            & BGE-base-en-v1.5& 8.314 \\ \midrule 
         \multirow{2}{*}{queryonly     }& BM25& 9.998 \\ 
                            & BGE-base-en-v1.5& 8.269 \\ \midrule 
         \multirow{2}{*}{doconly       }& BM25& 10.096 \\ 
                            & BGE-base-en-v1.5& 8.467 \\ \midrule 
         \multirow{2}{*}{Ours          }& BM25& \bf 9.501 \\ 
                            & BGE-base-en-v1.5& \bf 7.743 \\ \bottomrule 
    \end{tabular}}
    \label{tab:ablation_on_nfcorpus_ce}
    \end{minipage}
    \hfill
    \begin{minipage}{0.45\linewidth}
        \centering
        \captionof{table}{NDCG@10 Performance of Config Ablation Studies on NFCorpus.}
    \resizebox{1\linewidth}{!}{
        \begin{tabular}{clr}
        \toprule
        \multicolumn{2}{c}{Settings} & NDCG@10 \\ \midrule
         \multirow{2}{*}{Qwen2.5-7B    }& BM25& 0.363\\ 
                            & BGE-base-en-v1.5& 0.371\\ \midrule
        \multirow{2}{*}{groupnorm (GRPO)}& BM25& 0.376\\ 
                            & BGE-base-en-v1.5& 0.374\\ \midrule
        \multirow{2}{*}{batchnorm (RF++)}& BM25& 0.364\\ 
                            & BGE-base-en-v1.5& 0.354\\ \midrule
         \multirow{2}{*}{w/o adv scale }& BM25& 0.366\\ 
                            & BGE-base-en-v1.5& 0.348\\ \midrule
         \multirow{2}{*}{Ours          }& BM25& \bf 0.403\\ 
                            & BGE-base-en-v1.5& \bf 0.384\\ \bottomrule
    \end{tabular}}
    \label{tab:config_ablation_nfcorpus_norm}
    \end{minipage}
\end{table}

We conduct experiments to evaluate the effectiveness of our proposed reinforcement learning framework, as well as to investigate the importance of query-document collaborative training. The experimental results substantiate the core motivations underlying our training framework and demonstrate its performance advantages. Additionally, we perform ablation studies to examine the impact of different advantage calculation strategies on model performance.

\subsection{Experimental Setup}

We adopt the BEIR benchmark~\citep{thakur2021beir} and select three of its datasets for model training. These datasets are used to evaluate the benefits of in-domain training and to assess model generalization through cross-benchmark testing. Training LLMs using RL is computationally expensive. Due to resource constraints, training is limited to a maximum of 300 steps per dataset, and we choose to allocate our resources to experiments centered around one base model, Qwen2.5-7B~\citep{qwen2025qwen25technicalreport}, for a fair evaluation of improvements.

To investigate the impact of query-document collaborative training, we conduct ablation experiments on the NFCorpus dataset. Specifically, we train models and evaluate them alone or in combination under three settings: query-augmentation-only, document-augmentation-only, and query-document collaborative augmentation.
Furthermore, we compare the effects of different advantage calculation methods via an additional ablation study. We compare existing normalization strategies from GRPO and REINFORCE++, as well as our centralization-only setting. All experiments adopt NDCG@10 ~\citep{jarvelin2002cumulated}  as the evaluation metric.

Notably, our framework is designed to be orthogonal to the underlying retrieval architecture, supporting flexible integration with different retrieval modules. To demonstrate this, we evaluate our method in both sparse and dense retrieval settings. For sparse retrieval, we adopt BM25 as the base retriever, where the LLM is prompted to first summarize textual content and then generate query expansions or document keywords in the form of discrete words. For dense retrieval, we utilize BGE-base-en-v1.5, where the LLM produces sentence-level query expansions or document summaries. These augmented outputs are concatenated with the original contents. These distinct augmentation formats align with the underlying retrieval mechanisms: BM25 relies on matching of discrete words, whereas the BGE-base-en-v1.5 operates on holistic paragraph understanding and vectorization.

\subsection{Performance on Sparse and Dense Retrieval}
Tab.~\ref{tab:quantitative} reports the performance comparison of approaches across three datasets, with the best and top-three results denoted in bold and underline, respectively. 
Compared to baseline retrievers, our training method consistently achieves superior results across different settings, with notable improvements of 5\%–7\% in the sparse retrieval scenario on all the in-domain datasets. 

{\bf Cross-benchmark generalization}.
We conducted cross-benchmark validation on the models trained on three different datasets to evaluate the generalization capability of our method. As observed, in the sparse retrieval setting, our models demonstrated strong generalization ability: models trained on any single dataset achieved notable performance improvements over the untrained Qwen2.5-7B when applied to unseen domains. In contrast, under the dense retrieval setting, the generalization performance varied across domains. While improvements were observed on NFCorpus and SciFact, the performance on FiQA-2018 was inferior to that of Qwen2.5-7B.

\subsection{Essentialness of Query-Document Co-Augmentation}

Tab.~\ref{tab:ablation_on_nfcorpus_ndcg} shows ablation studies on query-document co-augmentation. Note that models trained with only query or document augmentation, whether used alone or in combination, exhibit a significant performance gap compared with our collaboratively trained model. Results show that using RL to enhance the base model in query rewriting leads to significant performance gains, which aligns with previous works~\citep{ma2023queryrewritingretrievalaugmentedlarge, Fan2024ASO, Zhao2024RetrievalAG, hsu2024grounding}, yet simply adding document augmentation on top of this does not consistently lead to further improvements and may even cause degradation.

This demonstrates that the success of our approach cannot be simply attributed to the introduction of document augmentation or to using RL to enhance document augmentation itself, as the superior performance gain only emerges with collaborative training for bidirectional augmentation.

To further investigate the detailed behavior of our explored collaborative bidirectional augmentation policy, we conduct qualitative and quantitative analyses of the augmented results. We first compute the word distributions of the augmented queries and documents across the entire dataset for all models used in the ablation study of Tab.~\ref{tab:ablation_on_nfcorpus_ndcg}, and calculate the cross-entropy $H(Q, D)$ of the query distribution relative to the document distribution, which quantitatively reflects the behavior of the output token space given the same input. As shown in Tab.~\ref{tab:ablation_on_nfcorpus_ce}, the cross-entropy for models trained with collaborative query-document augmentation is substantially lower than that of other models in both sparse and dense retrieval settings. This supports our motivation in enhancing both queries and documents: collaborative training enables the policy to learn to cooperate with itself to match queries and documents in the semantic domain, thereby improving retrieval effectiveness, which is only possible when trained in a bidirectional manner. With the policy understanding how to align the queries and documents instead of overfitting to the knowledge of a specific corpus, this explains the observed generalization across datasets, which can be essential when deployed to a new corpus. We also note that the generalization ability from the BM25 model is stronger than that of the policy explored using the dense retrieval model. This might be caused by dense retrievers developing implicit representations with domain preference~\citep{zhao2024dense}, which is also observed and reported in our concurrent work on RL-based query augmentation~\citep{jiang2025deepretrieval}.

\begin{figure}
    \centering
    \includegraphics[width=0.8\linewidth]{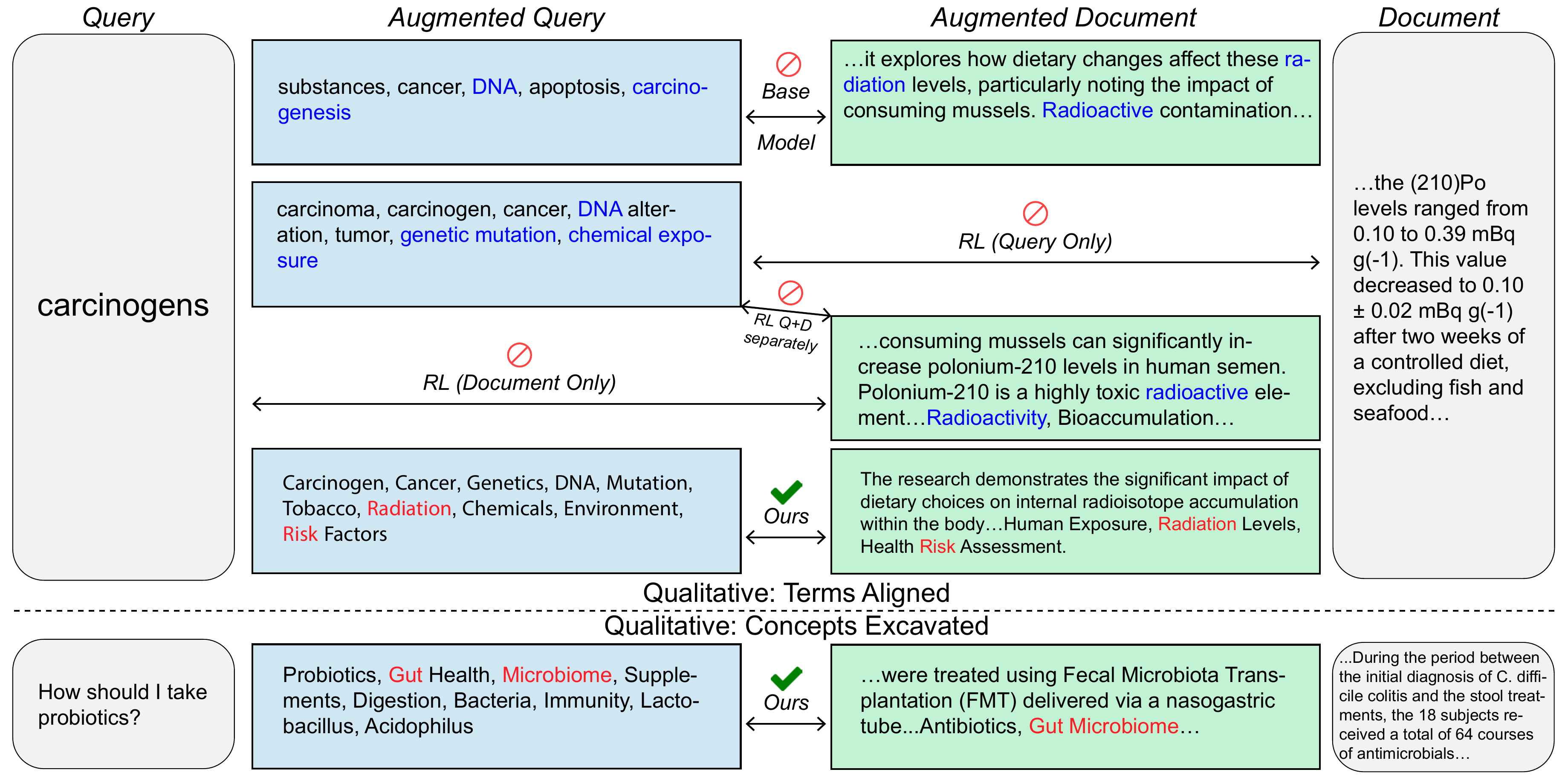}
    \caption{Qualitative demonstration of RL behaviors that boost performance. \textit{Top:} Concepts are discovered, but terms are not unified. Related terms are marked in \textcolor{blue}{blue}, while matched terms are marked in \textcolor{red}{red}. The base model, query-only, and document-only models all discovered related concepts, but with misaligned terms. Enhancing the documents alone does not directly improve performance; instead, success arises from joint bidirectional training. \textit{Below:} We show another case where our policy also learns to discover concepts to retrieve related documents, whereas other methods failed to generate related concepts (failed results omitted for conciseness).}
    \label{fig:qualitative_results}
\end{figure}

{\bf Case study}. We also provide qualitative analysis with a case study, focusing on analyzing sparse retrievers, as the retriever operates in word space, which is directly interpretable by humans, as presented in Fig.~\ref{fig:qualitative_results}. 
In the top case, we aimed to retrieve a target document concerning ``(210)Po'' using the query ``carcinogens''. It can be observed that all models were able to generate relevant words during the augmentation process that semantically ``bridge'' the gap between the query and the document. For example, words such as ``DNA'', ``carcinogenesis'', ``genetic mutation'', and ``chemical exposure'' were generated to augment the query, while words like ``radiation'' and ``radioactivity'' were generated to augment the document. However, these semantically related words failed to achieve a successful match due to discrepancies in word distribution. In contrast, our proposed method augments both the query and the document with a consistent word distribution, resulting in the presence of shared words such as ``radiation'' and ``risk'' in both augmented texts. This lexical alignment facilitates the successful retrieval of the target document.
In the bottom case, our method expands the query ``Probiotics'' with related words such as ``Gut Health'', ``Microbiome'', ``Supplements'', ``Digestion'', ``Bacteria'', ``Immunity'', ``Lactobacillus'', and ``Acidophilus'', and augments the document mentioning ``treated using Fecal Microbiota Transplantation (FMT)'' with words related to the ``Gut Microbiome'', thus successfully retrieving a target document with otherwise low textual overlap.

\subsection{Analysis on Normalization Techniques}
\label{subsec:ablation_studies}
Finally, we conduct ablation studies on the advantage calculation settings described in Sec.~\ref{taming_reward_var_in_rl} to highlight the performance benefits of our approach. Tab.~\ref{tab:config_ablation_nfcorpus_norm} compares the results of group normalization (GRPO), batch normalization (REINFORCE++), and centralization only (ours). Group normalization amplifies sampling variance, introducing excessive noise and impairing optimization. Batch normalization disrupts intra-group reward comparisons, causing advantage calculations within a batch to be dominated by query difficulty, resulting in negligible training effectiveness; in dense retrieval, it even underperforms direct augmentation with Qwen2.5-7B. In contrast, our centralization-only setting preserves intra-group reward relationships and thus achieves the best performance. Table 4 also demonstrates the necessity of advantage scaling after centralization: without it, the training direction is dominated by the majority of irrelevant documents, which constitute the majority within each batch, leading to poor performance.

\section{Limitations and Future Work}
\label{sec:limitation}
It is challenging to utilize our approach to conduct training when the number of candidate documents is very large. Even though this is not a significant issue during testing, since the augmentation of documents can be precomputed rather than generated online, the key techniques introduced in our approach are indeed for reducing the computational cost of generating significantly increased rollouts during training. This makes our approach more suitable for semantically complicated domains without large document sets, where retrieval performance can be boosted meanwhile computation cost remains no issue. We believe that this leads to important and meaningful future work in applying our approach to challenging IR applications. Furthermore, we believe that the bidirectional RL approach can also be of independent interest for other 
 RL scenarios where tackling the high-variance reward signals is among the key technical challenges.

\section*{Acknowledgement}
This work was supported by National Key R\&D Program of China (2023YFB3107102) and National Natural Science Foundation of China (62206245).

\bibliographystyle{named}
\bibliography{rag}

\begin{thebibliography}{}

\bibitem[\protect\citeauthoryear{Abdul-Jaleel \bgroup \em et al.\egroup }{2004}]{abdul2004umass}
Nasreen Abdul-Jaleel, James Allan, W~Bruce Croft, Fernando Diaz, Leah Larkey, Xiaoyan Li, Mark~D Smucker, and Courtney Wade.
\newblock Umass at trec 2004: Novelty and hard.
\newblock In {\em Proceedings of TREC-13}, pages 715--725, 2004.

\bibitem[\protect\citeauthoryear{Asai \bgroup \em et al.\egroup }{2024}]{asai2024selfrag}
Akari Asai, Zeqiu Wu, Yizhong Wang, Avirup Sil, and Hannaneh Hajishirzi.
\newblock Self-rag: Learning to retrieve, generate, and critique through self-reflection.
\newblock In {\em The Twelfth International Conference on Learning Representations}, 2024.

\bibitem[\protect\citeauthoryear{Baeza-Yates \bgroup \em et al.\egroup }{1999}]{baeza1999modern}
Ricardo Baeza-Yates, Berthier Ribeiro-Neto, et~al.
\newblock {\em Modern information retrieval}, volume 463.
\newblock ACM press New York, 1999.

\bibitem[\protect\citeauthoryear{Belkin \bgroup \em et al.\egroup }{1982}]{belkin1982ask}
Nicholas~J Belkin, Robert~N Oddy, and Helen~M Brooks.
\newblock Ask for information retrieval: Part i. background and theory.
\newblock {\em Journal of documentation}, 38(2):61--71, 1982.

\bibitem[\protect\citeauthoryear{Brown \bgroup \em et al.\egroup }{2020}]{brown2020language}
Tom Brown, Benjamin Mann, Nick Ryder, Melanie Subbiah, Jared~D Kaplan, Prafulla Dhariwal, Arvind Neelakantan, Pranav Shyam, Girish Sastry, Amanda Askell, et~al.
\newblock Language models are few-shot learners.
\newblock {\em Advances in neural information processing systems}, 33:1877--1901, 2020.

\bibitem[\protect\citeauthoryear{Chen \bgroup \em et al.\egroup }{2017}]{chen2017reading}
Danqi Chen, Adam Fisch, Jason Weston, and Antoine Bordes.
\newblock Reading wikipedia to answer open-domain questions.
\newblock In {\em Proceedings of the 55th Annual Meeting of the Association for Computational Linguistics (Volume 1: Long Papers)}, pages 1870--1879, 2017.

\bibitem[\protect\citeauthoryear{Christiano \bgroup \em et al.\egroup }{2017}]{Christiano2017DeepRL}
Paul~Francis Christiano, Jan Leike, Tom~B. Brown, Miljan Martic, Shane Legg, and Dario Amodei.
\newblock Deep reinforcement learning from human preferences.
\newblock {\em ArXiv}, abs/1706.03741, 2017.

\bibitem[\protect\citeauthoryear{Dageville \bgroup \em et al.\egroup }{2016}]{dageville2016snowflake}
Benoit Dageville, Thierry Cruanes, Marcin Zukowski, Vadim Antonov, Artin Avanes, Jon Bock, Jonathan Claybaugh, Daniel Engovatov, Martin Hentschel, Jiansheng Huang, et~al.
\newblock The snowflake elastic data warehouse.
\newblock In {\em Proceedings of the 2016 International Conference on Management of Data}, pages 215--226, 2016.

\bibitem[\protect\citeauthoryear{Dai \bgroup \em et al.\egroup }{2024a}]{dai2024cocktailcomprehensiveinformationretrieval}
Sunhao Dai, Weihao Liu, Yuqi Zhou, Liang Pang, Rongju Ruan, Gang Wang, Zhenhua Dong, Jun Xu, and Ji-Rong Wen.
\newblock Cocktail: A comprehensive information retrieval benchmark with llm-generated documents integration, 2024.

\bibitem[\protect\citeauthoryear{Dai \bgroup \em et al.\egroup }{2024b}]{dai2024bias}
Sunhao Dai, Chen Xu, Shicheng Xu, Liang Pang, Zhenhua Dong, and Jun Xu.
\newblock Bias and unfairness in information retrieval systems: New challenges in the llm era.
\newblock In {\em Proceedings of the 30th ACM SIGKDD Conference on Knowledge Discovery and Data Mining}, pages 6437--6447, 2024.

\bibitem[\protect\citeauthoryear{Dalton \bgroup \em et al.\egroup }{2014}]{Dalton2014EntityQF}
Jeffrey Dalton, Laura Dietz, and James Allan.
\newblock Entity query feature expansion using knowledge base links.
\newblock {\em Proceedings of the 37th international ACM SIGIR conference on Research \& development in information retrieval}, 2014.

\bibitem[\protect\citeauthoryear{Douze \bgroup \em et al.\egroup }{2024}]{douze2024faiss}
Matthijs Douze, Alexandr Guzhva, Chengqi Deng, Jeff Johnson, Gergely Szilvasy, Pierre-Emmanuel Mazar{\'e}, Maria Lomeli, Lucas Hosseini, and Herv{\'e} J{\'e}gou.
\newblock The faiss library.
\newblock {\em arXiv preprint arXiv:2401.08281}, 2024.

\bibitem[\protect\citeauthoryear{Fan \bgroup \em et al.\egroup }{2024}]{Fan2024ASO}
Wenqi Fan, Yujuan Ding, Liang bo~Ning, Shijie Wang, Hengyun Li, Dawei Yin, Tat-Seng Chua, and Qing Li.
\newblock A survey on rag meeting llms: Towards retrieval-augmented large language models.
\newblock In {\em Knowledge Discovery and Data Mining}, 2024.

\bibitem[\protect\citeauthoryear{Gao \bgroup \em et al.\egroup }{2023}]{gao2023retrieval}
Yunfan Gao, Yun Xiong, Xinyu Gao, Kangxiang Jia, Jinliu Pan, Yuxi Bi, Yixin Dai, Jiawei Sun, Haofen Wang, and Haofen Wang.
\newblock Retrieval-augmented generation for large language models: A survey.
\newblock {\em arXiv preprint arXiv:2312.10997}, 2:1, 2023.

\bibitem[\protect\citeauthoryear{Guo \bgroup \em et al.\egroup }{2025}]{deepseekai2025deepseekr1incentivizingreasoningcapability}
Daya Guo, Dejian Yang, Haowei Zhang, Junxiao Song, Ruoyu Zhang, Runxin Xu, Qihao Zhu, Shirong Ma, Peiyi Wang, Xiao Bi, et~al.
\newblock Deepseek-r1: Incentivizing reasoning capability in llms via reinforcement learning.
\newblock {\em arXiv preprint arXiv:2501.12948}, 2025.

\bibitem[\protect\citeauthoryear{Guu \bgroup \em et al.\egroup }{2020}]{guu2020retrieval}
Kelvin Guu, Kenton Lee, Zora Tung, Panupong Pasupat, and Mingwei Chang.
\newblock Retrieval augmented language model pre-training.
\newblock In {\em International conference on machine learning}, pages 3929--3938. PMLR, 2020.

\bibitem[\protect\citeauthoryear{Hsu \bgroup \em et al.\egroup }{2024}]{hsu2024grounding}
Sheryl Hsu, Omar Khattab, Chelsea Finn, and Archit Sharma.
\newblock Grounding by trying: Llms with reinforcement learning-enhanced retrieval.
\newblock {\em arXiv preprint arXiv:2410.23214}, 2024.

\bibitem[\protect\citeauthoryear{Hu}{2025a}]{Hu2025REINFORCEAS}
Jian Hu.
\newblock Reinforce++: A simple and efficient approach for aligning large language models.
\newblock {\em ArXiv}, abs/2501.03262, 2025.

\bibitem[\protect\citeauthoryear{Hu}{2025b}]{hu2025reinforce++}
Jian Hu.
\newblock Reinforce++: A simple and efficient approach for aligning large language models.
\newblock {\em arXiv preprint arXiv:2501.03262}, 2025.

\bibitem[\protect\citeauthoryear{J{\"a}rvelin and Kek{\"a}l{\"a}inen}{2002}]{jarvelin2002cumulated}
Kalervo J{\"a}rvelin and Jaana Kek{\"a}l{\"a}inen.
\newblock Cumulated gain-based evaluation of ir techniques.
\newblock {\em ACM Transactions on Information Systems (TOIS)}, 20(4):422--446, 2002.

\bibitem[\protect\citeauthoryear{Jiang \bgroup \em et al.\egroup }{2022}]{jiang-etal-2022-reatt}
Zhengbao Jiang, Luyu Gao, Jun Araki, Haibo Ding, Zhiruo Wang, Jamie Callan, and Graham Neubig.
\newblock Retrieval as attention: End-to-end learning of retrieval and reading within a single transformer.
\newblock In {\em Conference on Empirical Methods in Natural Language Processing (EMNLP)}, Abu Dhabi, UAE, December 2022.

\bibitem[\protect\citeauthoryear{Jiang \bgroup \em et al.\egroup }{2025}]{jiang2025deepretrieval}
Pengcheng Jiang, Jiacheng Lin, Lang Cao, Runchu Tian, SeongKu Kang, Zifeng Wang, Jimeng Sun, and Jiawei Han.
\newblock Deepretrieval: Hacking real search engines and retrievers with large language models via reinforcement learning.
\newblock {\em arXiv preprint arXiv:2503.00223}, 2025.

\bibitem[\protect\citeauthoryear{Karpukhin \bgroup \em et al.\egroup }{2020}]{karpukhin-etal-2020-dense}
Vladimir Karpukhin, Barlas Oguz, Sewon Min, Patrick Lewis, Ledell Wu, Sergey Edunov, Danqi Chen, and Wen-tau Yih.
\newblock Dense passage retrieval for open-domain question answering.
\newblock In {\em Proceedings of the 2020 Conference on Empirical Methods in Natural Language Processing (EMNLP)}, pages 6769--6781, Online, November 2020. Association for Computational Linguistics.

\bibitem[\protect\citeauthoryear{Khattab \bgroup \em et al.\egroup }{2022}]{khattab2022demonstrate}
Omar Khattab, Keshav Santhanam, Xiang~Lisa Li, David Hall, Percy Liang, Christopher Potts, and Matei Zaharia.
\newblock Demonstrate-search-predict: Composing retrieval and language models for knowledge-intensive nlp, 2022.
\newblock arXiv preprint arXiv:2212.14024.

\bibitem[\protect\citeauthoryear{Kwon \bgroup \em et al.\egroup }{2023}]{kwon2023efficient}
Woosuk Kwon, Zhuohan Li, Siyuan Zhuang, Ying Sheng, Lianmin Zheng, Cody~Hao Yu, Joseph~E. Gonzalez, Hao Zhang, and Ion Stoica.
\newblock Efficient memory management for large language model serving with pagedattention.
\newblock In {\em Proceedings of the ACM SIGOPS 29th Symposium on Operating Systems Principles}, 2023.

\bibitem[\protect\citeauthoryear{Lazaridou \bgroup \em et al.\egroup }{2022}]{lazaridou2022internet}
Angeliki Lazaridou, Elena Gribovskaya, Wojciech Stokowiec, and Nikolai Grigorev.
\newblock Internet-augmented language models through few-shot prompting for open-domain question answering.
\newblock {\em arXiv preprint arXiv:2203.05115}, 2022.

\bibitem[\protect\citeauthoryear{Lee \bgroup \em et al.\egroup }{2019}]{lee2019latent}
Kenton Lee, Ming-Wei Chang, and Kristina Toutanova.
\newblock Latent retrieval for weakly supervised open domain question answering.
\newblock In {\em Proceedings of the 57th Annual Meeting of the Association for Computational Linguistics}, pages 6086--6096, 2019.

\bibitem[\protect\citeauthoryear{Lee \bgroup \em et al.\egroup }{2022}]{lee-etal-2022-need}
Haejun Lee, Akhil Kedia, Jongwon Lee, Ashwin Paranjape, Christopher Manning, and Kyoung-Gu Woo.
\newblock You only need one model for open-domain question answering.
\newblock In {\em Proceedings of the 2022 Conference on Empirical Methods in Natural Language Processing}, pages 3047--3060, Abu Dhabi, United Arab Emirates, December 2022. Association for Computational Linguistics.

\bibitem[\protect\citeauthoryear{Lewis \bgroup \em et al.\egroup }{2020}]{lewis2020rag}
Patrick Lewis, Ethan Perez, Aleksandra Piktus, Fabio Petroni, Vladimir Karpukhin, Naman Goyal, Heinrich K{\"u}ttler, Mike Lewis, Wen-tau Yih, Tim Rockt{\"a}schel, et~al.
\newblock Retrieval-augmented generation for knowledge-intensive nlp tasks.
\newblock {\em Advances in Neural Information Processing Systems}, 33:9459--9474, 2020.

\bibitem[\protect\citeauthoryear{Ma \bgroup \em et al.\egroup }{2023a}]{ma2023queryrewritingretrievalaugmentedlarge}
Xinbei Ma, Yeyun Gong, Pengcheng He, Hai Zhao, and Nan Duan.
\newblock Query rewriting for retrieval-augmented large language models, 2023.

\bibitem[\protect\citeauthoryear{Ma \bgroup \em et al.\egroup }{2023b}]{ma2023query}
Xinbei Ma, Yeyun Gong, Pengcheng He, Hai Zhao, and Nan Duan.
\newblock Query rewriting in retrieval-augmented large language models.
\newblock In {\em Proceedings of the 2023 Conference on Empirical Methods in Natural Language Processing}, pages 5303--5315, 2023.

\bibitem[\protect\citeauthoryear{Mallen \bgroup \em et al.\egroup }{2022}]{mallen2023llm_memorization}
Alex Mallen, Akari Asai, Victor Zhong, Rajarshi Das, Hannaneh Hajishirzi, and Daniel Khashabi.
\newblock When not to trust language models: Investigating effectiveness and limitations of parametric and non-parametric memories.
\newblock {\em arXiv preprint}, 2022.

\bibitem[\protect\citeauthoryear{Mao \bgroup \em et al.\egroup }{2024}]{mao2024rafe}
Shengyu Mao, Yong Jiang, Boli Chen, Xiao Li, Peng Wang, Xinyu Wang, Pengjun Xie, Fei Huang, Huajun Chen, and Ningyu Zhang.
\newblock Rafe: ranking feedback improves query rewriting for rag.
\newblock {\em arXiv preprint arXiv:2405.14431}, 2024.

\bibitem[\protect\citeauthoryear{Meij \bgroup \em et al.\egroup }{2010}]{Meij2010ConceptualLM}
Edgar Meij, Dolf Trieschnigg, M.~de~Rijke, and Wessel Kraaij.
\newblock Conceptual language models for domain-specific retrieval.
\newblock {\em Inf. Process. Manag.}, 46:448--469, 2010.

\bibitem[\protect\citeauthoryear{Metzler and Croft}{2005}]{metzler2005markov}
Donald Metzler and W~Bruce Croft.
\newblock A markov random field model for term dependencies.
\newblock In {\em Proceedings of the 28th annual international ACM SIGIR conference on Research and development in information retrieval}, pages 472--479, 2005.

\bibitem[\protect\citeauthoryear{Metzler and Croft}{2007}]{metzler2007latent}
Donald Metzler and W~Bruce Croft.
\newblock Latent concept expansion using markov random fields.
\newblock In {\em Proceedings of the 30th annual international ACM SIGIR conference on Research and development in information retrieval}, pages 311--318, 2007.

\bibitem[\protect\citeauthoryear{Moritz \bgroup \em et al.\egroup }{2018}]{moritz2018raydistributedframeworkemerging}
Philipp Moritz, Robert Nishihara, Stephanie Wang, Alexey Tumanov, Richard Liaw, Eric Liang, Melih Elibol, Zongheng Yang, William Paul, Michael~I. Jordan, and Ion Stoica.
\newblock Ray: A distributed framework for emerging ai applications, 2018.

\bibitem[\protect\citeauthoryear{Ouyang \bgroup \em et al.\egroup }{2022}]{Ouyang2022TrainingLM}
Long Ouyang, Jeff Wu, Xu~Jiang, Diogo Almeida, Carroll~L. Wainwright, Pamela Mishkin, Chong Zhang, Sandhini Agarwal, Katarina Slama, Alex Ray, John Schulman, Jacob Hilton, Fraser Kelton, Luke~E. Miller, Maddie Simens, Amanda Askell, Peter Welinder, Paul~Francis Christiano, Jan Leike, and Ryan~J. Lowe.
\newblock Training language models to follow instructions with human feedback.
\newblock {\em ArXiv}, abs/2203.02155, 2022.

\bibitem[\protect\citeauthoryear{Pan \bgroup \em et al.\egroup }{2025}]{tinyzero}
Jiayi Pan, Junjie Zhang, Xingyao Wang, Lifan Yuan, Hao Peng, and Alane Suhr.
\newblock Tinyzero.
\newblock https://github.com/Jiayi-Pan/TinyZero, 2025.
\newblock Accessed: 2025-01-24.

\bibitem[\protect\citeauthoryear{Press \bgroup \em et al.\egroup }{2022}]{press2022measuring}
Ofir Press, Muru Zhang, Sewon Min, Ludwig Schmidt, Noah~A Smith, and Mike Lewis.
\newblock Measuring and narrowing the compositionality gap in language models.
\newblock {\em arXiv preprint arXiv:2210.03350}, 2022.

\bibitem[\protect\citeauthoryear{Qwen \bgroup \em et al.\egroup }{2025}]{qwen2025qwen25technicalreport}
Qwen, :, An~Yang, Baosong Yang, Beichen Zhang, Binyuan Hui, Bo~Zheng, Bowen Yu, Chengyuan Li, Dayiheng Liu, Fei Huang, Haoran Wei, Huan Lin, Jian Yang, Jianhong Tu, Jianwei Zhang, Jianxin Yang, Jiaxi Yang, Jingren Zhou, Junyang Lin, Kai Dang, Keming Lu, Keqin Bao, Kexin Yang, Le~Yu, Mei Li, Mingfeng Xue, Pei Zhang, Qin Zhu, Rui Men, Runji Lin, Tianhao Li, Tianyi Tang, Tingyu Xia, Xingzhang Ren, Xuancheng Ren, Yang Fan, Yang Su, Yichang Zhang, Yu~Wan, Yuqiong Liu, Zeyu Cui, Zhenru Zhang, and Zihan Qiu.
\newblock Qwen2.5 technical report, 2025.

\bibitem[\protect\citeauthoryear{Robertson \bgroup \em et al.\egroup }{2009}]{robertson2009probabilistic}
Stephen Robertson, Hugo Zaragoza, et~al.
\newblock The probabilistic relevance framework: Bm25 and beyond.
\newblock {\em Foundations and Trends{\textregistered} in Information Retrieval}, 3(4):333--389, 2009.

\bibitem[\protect\citeauthoryear{Sachan \bgroup \em et al.\egroup }{2021}]{emdr22021SachanRHDY21}
Devendra~Singh Sachan, Siva Reddy, William~L. Hamilton, Chris Dyer, and Dani Yogatama.
\newblock End-to-end training of multi-document reader and retriever for open-domain question answering.
\newblock In Marc'Aurelio Ranzato, Alina Beygelzimer, Yann~N. Dauphin, Percy Liang, and Jennifer~Wortman Vaughan, editors, {\em Advances in Neural Information Processing Systems 34: Annual Conference on Neural Information Processing Systems 2021, NeurIPS 2021, December 6-14, 2021, virtual}, pages 25968--25981, 2021.

\bibitem[\protect\citeauthoryear{Salton and Buckley}{1988}]{salton1988term}
Gerard Salton and Christopher Buckley.
\newblock Term-weighting approaches in automatic text retrieval.
\newblock {\em Information processing \& management}, 24(5):513--523, 1988.

\bibitem[\protect\citeauthoryear{Schulman \bgroup \em et al.\egroup }{2017}]{Schulman2017ProximalPO}
John Schulman, Filip Wolski, Prafulla Dhariwal, Alec Radford, and Oleg Klimov.
\newblock Proximal policy optimization algorithms.
\newblock {\em ArXiv}, abs/1707.06347, 2017.

\bibitem[\protect\citeauthoryear{Shen \bgroup \em et al.\egroup }{2023}]{shen2023large}
Tao Shen, Guodong Long, Xiubo Geng, Chongyang Tao, Tianyi Zhou, and Daxin Jiang.
\newblock Large language models are strong zero-shot retriever.
\newblock {\em arXiv preprint arXiv:2304.14233}, 2023.

\bibitem[\protect\citeauthoryear{Sheng \bgroup \em et al.\egroup }{2024}]{sheng2024hybridflow}
Guangming Sheng, Chi Zhang, Zilingfeng Ye, Xibin Wu, Wang Zhang, Ru~Zhang, Yanghua Peng, Haibin Lin, and Chuan Wu.
\newblock Hybridflow: A flexible and efficient rlhf framework.
\newblock {\em arXiv preprint arXiv: 2409.19256}, 2024.

\bibitem[\protect\citeauthoryear{Shi \bgroup \em et al.\egroup }{2023}]{shi2023replug}
Weijia Shi, Sewon Min, Michihiro Yasunaga, Minjoon Seo, Rich James, Mike Lewis, Luke Zettlemoyer, and Wen-tau Yih.
\newblock Replug: Retrieval-augmented black-box language models.
\newblock {\em arXiv preprint arXiv:2301.12652}, 2023.

\bibitem[\protect\citeauthoryear{Singhal and others}{2001}]{singhal2001modern}
Amit Singhal et~al.
\newblock Modern information retrieval: A brief overview.
\newblock {\em IEEE Data Eng. Bull.}, 24(4):35--43, 2001.

\bibitem[\protect\citeauthoryear{Stiennon \bgroup \em et al.\egroup }{2020}]{Stiennon2020LearningTS}
Nisan Stiennon, Long Ouyang, Jeff Wu, Daniel~M. Ziegler, Ryan~J. Lowe, Chelsea Voss, Alec Radford, Dario Amodei, and Paul Christiano.
\newblock Learning to summarize from human feedback.
\newblock {\em ArXiv}, abs/2009.01325, 2020.

\bibitem[\protect\citeauthoryear{Syed}{2010}]{syed2010wikitology}
Zareen~Saba Syed.
\newblock {\em Wikitology: A novel hybrid knowledge base derived from wikipedia}.
\newblock University of Maryland, Baltimore County, 2010.

\bibitem[\protect\citeauthoryear{Thakur \bgroup \em et al.\egroup }{2021}]{thakur2021beir}
Nandan Thakur, Nils Reimers, Andreas R{\"u}ckl{\'e}, Abhishek Srivastava, and Iryna Gurevych.
\newblock Beir: A heterogenous benchmark for zero-shot evaluation of information retrieval models.
\newblock {\em arXiv preprint arXiv:2104.08663}, 2021.

\bibitem[\protect\citeauthoryear{Touvron \bgroup \em et al.\egroup }{2023}]{Touvron2023Llama2O}
Hugo Touvron, Louis Martin, Kevin~R. Stone, Peter Albert, Amjad Almahairi, Yasmine Babaei, Nikolay Bashlykov, Soumya Batra, Prajjwal Bhargava, Shruti Bhosale, et~al.
\newblock Llama 2: Open foundation and fine-tuned chat models.
\newblock {\em ArXiv}, abs/2307.09288, 2023.

\bibitem[\protect\citeauthoryear{Trivedi \bgroup \em et al.\egroup }{2023}]{trivedi2023interleavingretrievalchainofthoughtreasoning}
Harsh Trivedi, Niranjan Balasubramanian, Tushar Khot, and Ashish Sabharwal.
\newblock Interleaving retrieval with chain-of-thought reasoning for knowledge-intensive multi-step questions, 2023.

\bibitem[\protect\citeauthoryear{Wang \bgroup \em et al.\egroup }{2023}]{wang2023query2doc}
Liang Wang, Nan Yang, and Furu Wei.
\newblock Query2doc: Query expansion with large language models.
\newblock {\em arXiv preprint arXiv:2303.07678}, 2023.

\bibitem[\protect\citeauthoryear{Wang \bgroup \em et al.\egroup }{2024}]{wang2024multilingual}
Liang Wang, Nan Yang, Xiaolong Huang, Linjun Yang, Rangan Majumder, and Furu Wei.
\newblock Multilingual e5 text embeddings: A technical report.
\newblock {\em arXiv preprint arXiv:2402.05672}, 2024.

\bibitem[\protect\citeauthoryear{Xiao \bgroup \em et al.\egroup }{2024}]{xiao2024c}
Shitao Xiao, Zheng Liu, Peitian Zhang, Niklas Muennighoff, Defu Lian, and Jian-Yun Nie.
\newblock C-pack: Packed resources for general chinese embeddings.
\newblock In {\em Proceedings of the 47th international ACM SIGIR conference on research and development in information retrieval}, pages 641--649, 2024.

\bibitem[\protect\citeauthoryear{Xiong and Callan}{2015}]{Xiong2015QueryEW}
Chenyan Xiong and Jamie Callan.
\newblock Query expansion with freebase.
\newblock {\em Proceedings of the 2015 International Conference on The Theory of Information Retrieval}, 2015.

\bibitem[\protect\citeauthoryear{Xu \bgroup \em et al.\egroup }{2009}]{Xu2009QueryDP}
Yang Xu, G.~Jones, and Bin Wang.
\newblock Query dependent pseudo-relevance feedback based on wikipedia.
\newblock {\em Proceedings of the 32nd international ACM SIGIR conference on Research and development in information retrieval}, 2009.

\bibitem[\protect\citeauthoryear{Yao \bgroup \em et al.\egroup }{2023}]{yao2023reactsynergizingreasoningacting}
Shunyu Yao, Jeffrey Zhao, Dian Yu, Nan Du, Izhak Shafran, Karthik Narasimhan, and Yuan Cao.
\newblock React: Synergizing reasoning and acting in language models, 2023.

\bibitem[\protect\citeauthoryear{Ye \bgroup \em et al.\egroup }{2023}]{ye2023enhancing}
Fanghua Ye, Meng Fang, Shenghui Li, and Emine Yilmaz.
\newblock Enhancing conversational search: Large language model-aided informative query rewriting.
\newblock {\em arXiv preprint arXiv:2310.09716}, 2023.

\bibitem[\protect\citeauthoryear{Yu \bgroup \em et al.\egroup }{2025}]{yu2025dapoopensourcellmreinforcement}
Qiying Yu, Zheng Zhang, Ruofei Zhu, Yufeng Yuan, Xiaochen Zuo, Yu~Yue, Weinan Dai, Tiantian Fan, Gaohong Liu, Lingjun Liu, Xin Liu, Haibin Lin, Zhiqi Lin, Bole Ma, Guangming Sheng, Yuxuan Tong, Chi Zhang, Mofan Zhang, Wang Zhang, Hang Zhu, Jinhua Zhu, Jiaze Chen, Jiangjie Chen, Chengyi Wang, Hongli Yu, Yuxuan Song, Xiangpeng Wei, Hao Zhou, Jingjing Liu, Wei-Ying Ma, Ya-Qin Zhang, Lin Yan, Mu~Qiao, Yonghui Wu, and Mingxuan Wang.
\newblock Dapo: An open-source llm reinforcement learning system at scale, 2025.

\bibitem[\protect\citeauthoryear{Zesch \bgroup \em et al.\egroup }{2007}]{zesch2007analyzing}
Torsten Zesch, Iryna Gurevych, and Max M{\"u}hlh{\"a}user.
\newblock Analyzing and accessing wikipedia as a lexical semantic resource.
\newblock {\em Data Structures for Linguistic Resources and Applications}, 197205, 2007.

\bibitem[\protect\citeauthoryear{Zesch \bgroup \em et al.\egroup }{2008}]{zesch-etal-2008-extracting}
Torsten Zesch, Christof M{\"u}ller, and Iryna Gurevych.
\newblock Extracting lexical semantic knowledge from {W}ikipedia and {W}iktionary.
\newblock In Nicoletta Calzolari, Khalid Choukri, Bente Maegaard, Joseph Mariani, Jan Odijk, Stelios Piperidis, and Daniel Tapias, editors, {\em Proceedings of the Sixth International Conference on Language Resources and Evaluation ({LREC}`08)}, Marrakech, Morocco, May 2008. European Language Resources Association (ELRA).

\bibitem[\protect\citeauthoryear{Zhao \bgroup \em et al.\egroup }{2024a}]{Zhao2024RetrievalAG}
Siyun Zhao, Yuqing Yang, Zilong Wang, Zhiyuan He, Luna~K. Qiu, and Lili Qiu.
\newblock Retrieval augmented generation (rag) and beyond: A comprehensive survey on how to make your llms use external data more wisely.
\newblock {\em ArXiv}, abs/2409.14924, 2024.

\bibitem[\protect\citeauthoryear{Zhao \bgroup \em et al.\egroup }{2024b}]{zhao2024dense}
Wayne~Xin Zhao, Jing Liu, Ruiyang Ren, and Ji-Rong Wen.
\newblock Dense text retrieval based on pretrained language models: A survey.
\newblock {\em ACM Transactions on Information Systems}, 42(4):1--60, 2024.

\end{thebibliography}

\clearpage
\appendix

\section{Additional Experimental Results}
\subsection{Results on More Datasets}

\begin{table*}[t]
    \centering
        \caption{NDCG@10 Performance of models trained from Qwen2.5-7B on additional test-only datasets. The best results are bolded, and other top-three results are underlined.}
    \resizebox{.6\linewidth}{!}{
        \begin{tabular}{cccc}
        \toprule
        \multicolumn{2}{c}{Settings} & SCIDOCS & TREC-COVID \\ \midrule

         \multirow{2}{*}{Base Retriever}& BM25& 0.165& 0.688 \\ 
                            & BGE-base-en-v1.5& 0.214& 0.672 \\ \midrule
         \multirow{2}{*}{Qwen2.5-7B    }& BM25& 0.162& \underline{0.694} \\ 
                            & BGE-base-en-v1.5& 0.212& \underline{0.776} \\ \midrule
         \multirow{2}{*}{Ours-NF       }& BM25& \underline{0.173}& \underline{0.726} \\ 
                            & BGE-base-en-v1.5& \textbf{0.224}& \textbf{0.807} \\ \midrule
         \multirow{2}{*}{Ours-SF       }& BM25& \underline{0.168}& 0.672 \\ 
                            & BGE-base-en-v1.5& \underline{0.220}& \underline{0.790} \\ \midrule
         \multirow{2}{*}{Ours-FQ       }& BM25& \textbf{0.181}& \textbf{0.727} \\ 
                            & BGE-base-en-v1.5& \underline{0.218}& 0.759 \\ 
\bottomrule
    \end{tabular}}
    \label{tab:quantitative_test-only}
\end{table*}

To further validate the generalization performance, we conduct additional experiments on two datasets SCIDOCS, TREC-COVID from BEIR that contain only test sets. We evaluate both the base model Qwen2.5-7B and our three trained models on these test sets, with the results presented in Tab.~\ref{tab:quantitative_test-only}. The experimental results demonstrate that our models, trained in other domains, maintain significant performance improvements when transferred to these datasets. Specifically, in dense retrieval settings on the TREC-COVID dataset, collaborative augmentation from our model shows a 13\% improvement over the base retriever and a 3.1\% improvement over the augmentation from the base model.
When comparing our models trained on different datasets, we observe that Ours-NF and Ours-FQ slightly outperform Ours-SF. This performance gap likely stems from the limited size of the SciFact training set, which contains fewer than 1,000 query-document pairs, potentially restricting the model's learning capacity.

\subsection{Results of More Baselines}

\begin{table*}[t]
    \centering
        \caption{NDCG@10 Performance for more baselines. The best results are bolded, and other top-three results are underlined.}
    \resizebox{.8\linewidth}{!}{
        \begin{tabular}{ccccccc}
        \toprule
\multicolumn{2}{c}{Settings} & NFCorpus & SciFact & FiQA-2018 & SCIDOCS & TREC-COVID \\ \midrule

\multirow{2}{*}{Base Retriever}& BM25& 0.343& 0.691& 0.254& 0.165& 0.688 \\ 
                            & BGE-base-en-v1.5& 0.368& 0.738& \underline{0.391}& 0.214& 0.672 \\ \midrule
\multirow{2}{*}{Qwen2.5-3B    }& BM25& 0.352& 0.692& 0.237& 0.155& 0.642 \\ 
                            & BGE-base-en-v1.5& 0.363& 0.741& 0.359& 0.210& \underline{0.772} \\ \midrule
\multirow{2}{*}{Qwen2.5-7B    }& BM25& 0.363& 0.696& 0.258& 0.162& \underline{0.694} \\ 
                            & BGE-base-en-v1.5& 0.371& 0.746& 0.383& 0.212& \underline{0.776} \\ \midrule
\multirow{2}{*}{Qwen2.5-72B   }& BM25& \underline{0.370}& \underline{0.747}& \underline{0.305}& \underline{0.164}& 0.670 \\ 
                            & BGE-base-en-v1.5& \underline{0.377}& \underline{0.751}& \textbf{0.395}& \underline{0.220}& 0.734 \\ \midrule
\multirow{2}{*}{Ours-3B       }& BM25& \underline{0.371}& \underline{0.715}& \underline{0.273}& \underline{0.168}& \underline{0.716} \\ 
                            & BGE-base-en-v1.5& \underline{0.379}& \underline{0.749}& 0.364& \underline{0.217}& 0.771 \\ \midrule
\multirow{2}{*}{Ours-7B       }& BM25& \textbf{0.403}& \textbf{0.748}& \textbf{0.328}& \textbf{0.181}& \textbf{0.727} \\ 
                            & BGE-base-en-v1.5& \textbf{0.384}& \textbf{0.753}& \textbf{0.395}& \textbf{0.224}& \textbf{0.803} \\ 

\bottomrule
    \end{tabular}}
    \label{tab:quantitative_base-model}
\end{table*}

We further incorporate additional baselines for evaluation, including Qwen2.5-3B, Qwen2.5-7B and Qwen2.5-72B, to demonstrate the benefits of query-document correlative training. Our training framework is also applied to Qwen2.5-3B. Limited by computational resources, we did not use our method to train under Qwen2.5-72B. Tab.~\ref{tab:quantitative_base-model} presents the performance of baselines under both sparse and dense settings, where ``ours-3B'' and ``ours-7B'' denote the best-performing models trained on Qwen2.5-3B and Qwen2.5-7B, respectively.

We observe that Ours-7B achieves the best results across all datasets in both the sparse setting and the dense setting, surpassing Qwen2.5-72B, which has a substantially larger number of parameters. For Ours-3B, we also observe performance improvements in most settings compared to the original Qwen2.5-3B before training, with an increase of up to 7.4\% on the TREC-COVID dataset in the sparse setting. These results indicate that our approach provides significant improvements for base models in different scales. However, compared to larger models such as Qwen2.5-72B, the performance of Ours-3B is less impressive. Upon examining the augmentation outputs, we find a considerable number of empty strings, excessively long outputs, and non-English characters. One possible explanation is that the instruction-following capability of the Qwen2.5-3B model is relatively weak, making it difficult to ensure that the augmentation outputs are generated in the correct format, thereby limiting the model’s improvements through training.

We also observe that, for training-free models, performance generally correlates positively with model size, though this relationship is not strictly monotonic. In some cases, larger models produce less effective augmentation results, or the augmentation outputs perform worse than Base Retriever. This observation aligns with our expectations: for training-free models, queries and documents are augmented as independent text segments, resulting in output spaces that are possibly misaligned. Although the augmentation outputs for both queries and documents may be sufficiently comprehensive, their lack of alignment offers no benefit to retrieval performance.

\subsection{Results of Statistical Evaluation on Normalization Techniques}
\label{sec:analysis_anormaly}

\begin{table*}[t]
    \centering
        \caption{Anomalous group proportion under different normalization settings. ``Amplified Variance'' means the variance in reward is significantly amplified after normalization. ``Same Sign'' means all advantages in the group are positive or negative. ``*'' indicates this problem will not occur in theory. For more details, please see Sec.~\ref{sec:analysis_anormaly}}
    \resizebox{.6\linewidth}{!}{
        \begin{tabular}{cccc}
        \toprule
        \multicolumn{2}{c}{Settings} & Amplified Variance & Same Sign \\ \midrule

         \multirow{2}{*}{groupnorm (GRPO)}& BM25 & 94.4\%& 0.0\%* \\ 
                              & BGE-base-en-v1.5 & 94.8\%& 0.0\%* \\ \midrule
         \multirow{2}{*}{batchnorm (RF++)}& BM25 & 19.1\%& 63.9\% \\ 
                              & BGE-base-en-v1.5 & 7.5\%& 84.4\% \\ \midrule
         \multirow{2}{*}{Ours            }& BM25 & 0.0\%*& 0.0\%* \\ 
                              & BGE-base-en-v1.5 & 0.0\%*& 0.0\%* \\ 
\bottomrule
    \end{tabular}}
    \label{tab:anomalous_group_ratio}
\end{table*}

In this section, in addition to the empirical evidence from the ablation study, we also present statistical evidence to support the problem analysis in Sec.~\ref{taming_reward_var_in_rl} and Sec.~\ref{subsec:ablation_studies}. The experiments are conducted using the same settings as our ablation studies.

Specifically, we collect statistical evidence for the two problems illustrated in Fig.~\ref{fig:sampling_strategy}: \emph{Variance Amplified} and \emph{Biased Towards Easier Cases}. For a clear quantitative demonstration, we define two anomaly indicators and report their occurrences. To detect \emph{amplified variance}, we track the original percentage of groups with variance below a threshold $std_{threshold}$, and subtract the corresponding percentage after normalization. To detect \emph{biased advantage}, we evaluate the occurrence of all advantages in a group sharing the same sign (all positive or all negative), which strongly indicates bias introduced by batch normalization.

As shown in Tab.~\ref{tab:anomalous_group_ratio}, we observe that in our task, GRPO and our method do not exhibit the biased advantage reflected by the occurrence of the ``same sign'' phenomenon, as there is no batch-based advantage calculation. Regarding amplified variance, we choose a small threshold, and the resulting statistics demonstrate two key points: 1) there is initially a large proportion of rewards with small variance below our set threshold, and 2) these are amplified after normalization, especially in GRPO, as values below $0.02$ are unavoidably amplified to $1$. The selection of the threshold does not affect our conclusion, as it clearly shows that many rewards have very small variances, which are greatly amplified. We believe the abundance of small-variance groups is due to the fact that augmentation does not necessarily alter the similarity ranking in retrieval, especially for documents, where each batch contains many randomly irrelevant documents whose augmented results are unlikely to affect the reward. This also indicates that the variance introduced by our reward sampling computation is not significant, and that the problem is caused by subsequent computation rather than the variance of the reward sampling itself. While batch normalization reduces this problem, it causes many groups to have only positive or negative advantages, which does not represent a meaningful direction of optimization. In contrast, by only performing centering and allowing the variance to be naturally controlled by the reward computation itself, we avoid amplifying sampling errors in the reward and do not introduce intra-group bias, enabling effective reinforcement learning and successfully improving performance.

\section{Implementation Details}

\subsection{Training Setup}

We adopt a GRPO-based training pipeline, following most hyperparameters from the countdown experiment in TinyZero~\citep{tinyzero}, with several modifications to accommodate the requirements of different tasks. Specifically, the temperature is set to 1.2 to encourage exploration of diverse augmentations. A repetition penalty of 1.2 is applied to prevent the model from becoming trapped in locally optimal solutions that simply copy the original text. The micro batch size is set to 16, with each micro batch containing one query and 1–5 relevant documents (depending on the dataset), while the remaining slots are filled with randomly selected irrelevant documents. The batch size is set to 512 and the mini batch size to 128 to ensure gradient stability. We select advantage scale coefficients of 1.0, 0.2, and 0.1 for queries, relevant documents, and irrelevant documents, respectively, to balance the proportions of each component. Additionally, we remove the KL loss and entropy loss terms, as these constraints hinder further model optimization, which is consistent with findings reported in concurrent work DAPO~\citep{yu2025dapoopensourcellmreinforcement}. The format reward is also omitted; however, we still extract augmented content from within <answer></answer> tags, defaulting to empty augmentation if such tags are absent. Under this configuration, the model quickly learns to follow the required format while avoiding performance degradation due to excessive adherence to formatting.

\subsection{Environment Setup}

We utilize VERL~\citep{sheng2024hybridflow} as the LLM RL fine-tuning framework, building upon TinyZero. The software packages and runtime environment are configured to be compatible with this version of the training framework, including Python (v3.9), CUDA (v12.4), VLLM (v0.6.3)~\citep{kwon2023efficient}, PyTorch (v2.4.0), and Ray (v2.46.0)~\citep{moritz2018raydistributedframeworkemerging}. For sparse retrieval and BM25 evaluation, we employ ElasticSearch (v7.10.2), while FAISS-GPU (v1.7.2)~\citep{douze2024faiss} is used to support efficient dense vector matching. All experiments are conducted on a single node equipped with eight NVIDIA A100 80GB GPUs.

\end{document}